\documentclass[9pt,twocolumn,twoside]{pnas-new-custom}

\usepackage{multirow}
\usepackage{hyperref}
\hypersetup{
    colorlinks=true,
    linkcolor=blue,
    filecolor=magenta,      
    urlcolor=blue,
}

\templatetype{pnasresearcharticle} 
\title{New HST data and modeling reveal a massive planetesimal collision around Fomalhaut}

\author[a,1]{Andr\'as G\'asp\'ar} 
\author[a]{George H.\ Rieke} 
\affil[a]{Steward Observatory, The University of Arizona, 933 N Cherry Ave, Tucson, AZ, 85719}
\leadauthor{G\'asp\'ar} 

\significancestatement{
Although originally thought to be a massive exoplanet, the faintness of Fomalhaut b in the infrared and its failure to perturb Fomalhaut’s debris ring indicate a low mass. We use all available data to reveal that it has faded in brightness and grown in extent, with motion consistent with an escaping orbit. This behavior confirms suggestions that the source is a dispersing cloud of dust, produced by a massive collision between two planetesimals. The visible signature appears to be very fine dust escaping under the influence of radiation pressure. Such events should be rare in quiescent planetary systems at the age of Fomalhaut, suggesting increased dynamical activity within the system possibly due to orbital migration of hypothetical planets.
}

\authorcontributions{Author contributions: A.G. and G.H.R. designed research; A.G. performed research; A.G. contributed analytic tools; A.G. analyzed data; and A.G. and G.H.R. wrote the paper.}
\authordeclaration{The authors declare no competing interest.\\\\Original version of this article is a PNAS Direct Submission.\\\\Data deposition: The raw observational data can be downloaded from the Mikulski Archive for Space Telescopes (MAST) website maintained by the Space Telescope Science Institute (STScI). Our modeling code, DiskDyn, can be downloaded from \href{https://github.com/merope82/DiskDyn}{https://github.com/merope82/DiskDyn}.}
\correspondingauthor{\textsuperscript{1}To whom correspondence should be addressed. E-mail: agaspar@arizona.edu}


\begin{abstract}
The apparent detection of an exoplanet orbiting Fomalhaut was announced in 2008. However, subsequent observations of Fomalhaut b raised questions about its status: Unlike other exoplanets, it is bright in the optical and nondetected in the infrared, and its orbit appears to cross the debris ring around the star without the expected gravitational perturbations. We revisit previously published data and analyze additional Hubble Space Telescope (HST) data, finding that the source is likely on a radial trajectory and has faded and become extended. Dynamical and collisional modeling of a recently produced dust cloud yields results consistent with the observations. Fomalhaut b appears to be a directly imaged catastrophic collision between two large planetesimals in an extrasolar planetary system. Similar events should be very rare in quiescent planetary systems of the age of Fomalhaut, suggesting that we are possibly witnessing the effects of gravitational stirring due to the orbital evolution of hypothetical planet(s) around the star.
\end{abstract}

\dates{Approved January 15, 2020 (received for review July 19, 2019). arXiv compiled: \today. NOT ACTUAL PNAS ARTICLE}
\doi{Original Article at: \href{www.pnas.org/cgi/doi/10.1073/pnas.1912506117}{www.pnas.org/cgi/doi/10.1073/pnas.1912506117}}

\begin{document}

\maketitle
\thispagestyle{firststyle}
\ifthenelse{\boolean{shortarticle}}{\ifthenelse{\boolean{singlecolumn}}{\abscontentformatted}{\abscontent}}{}

\dropcap{T}he simultaneous announcements of images of massive planets around Fomalhaut \citep{kalas08} and 
HR 8799 \citep{marois08} were a benchmark in our exploration of exoplanets. The HR 8799 system has indeed 
become the prototype for complex systems of very massive planets and has been the subject of many studies of 
such objects \citep[e.g.,][]{marois10,barman11,currie11,marley12,zurlo16}. However, Fomalhaut b has been enigmatic. 

An issue with the massive planet hypothesis for Fomalhaut b was the nondetection with the Infrared Array Camera 
(IRAC) onboard the Spitzer Space Telescope \citep{marengo09}, which placed an upper limit of 3 Jupiter masses (M$_{\rm Jup}$) on its 
mass assuming an age for Fomalhaut of 200 Myr. A similar limit was derived from the lack of apparent perturbations 
in the Fomalhaut debris ring; this work favored a significantly smaller mass, e.g., 0.5 M$_{\rm Jup}$ \citep{chiang09}. An upward revision of 
the system age to 440 $\pm$ 40 Myr \citep{mamajek12} relaxed the limit from the infrared data, but a much deeper limit was obtained with 
Spitzer IRAC that pushed the mass limit lower. In fact, this limit appeared to be incompatible with a massive planet 
accounting for the visible brightness of the object \citep{janson12,janson15}. With determination of an orbit for Fomalhaut b, it became 
apparent that the object would cross the ring at least in projection \citep{kalas13}. The implications on the ring structure placed 
a tentative limit on the mass of Fomalhaut b as low as an Earth mass \citep{beust14}.

Given that the bright optical signature of Fomalhaut b seems not to be light scattered from a giant planet, a
number of alternative hypotheses have been proposed, e.g., light scattered by a circumplanetary ring system \citep{kalas08}
or by a dust cloud associated with a relatively low mass planet \citep[e.g.,][]{kennedy11,janson12,currie12}.
A tentative finding that the image of Fomalhaut b might be  extended was interpreted to support the dust cloud hypothesis
\citep{galicher13}; however, it was suggested that this effect instead might be due to speckle and other noise
sources \citep{kalas13,kenyon14}. Nonetheless, a dust cloud appears to be the most plausible
hypothesis.

A number of papers have addressed the origin of this hypothetical dust cloud \citep{kennedy11,galicher13,kenyon14,lawler15}.
\cite{kennedy11} suggested that collisions among a swarm of satellites around a planet could be responsible, and
showed that certain models of this process lie within the  constraints from assuming that the phenomenon has persisted
for the life of the star and that the image is point-like. \cite{kenyon15} found that satellite
swarms around planets of mass 10 -- 100 M$_{\rm Earth}$ that have evolved for 100 -- 400 Myr could match the properties of Fomalhaut b.
\cite{galicher13} suggested that the object could consist of dust created in the collision of two modest sized (50 km)
planetesimals similar to members of the Kuiper Belt. \cite{kenyon14}
analyzed three possible origins for the required dust cloud: 1) a giant planetesimal impact, 2) material captured from the
prominent debris disk of the star, or 3) dust generated in a collisional cascade from a massive cloud of satellites
around a recently formed planet. The first possibility was judged to be unlikely to reproduce the observations (see also \cite{tamayo14}).
The other two possibilities are constrained significantly by the assumption that the dust system should persist in a state similar
to its present one for the main sequence lifetime of
the star, i.e., $\sim 400$ Myr. \cite{lawler15} suggested a possible solution
to the lifetime issues by attributing the source to a transient dust cloud produced by a collision between planetesimals interior
to the main debris belt around the star.

We report that Fomalhaut b has grown in extent and faded since its discovery in Hubble Space Telescope ({\it HST}) images from 2004, 
with motion consistent with an escaping trajectory. This behavior is consistent with expectations for a dust cloud produced 
in a planetesimal collision and  dispersing dynamically. As the cloud disperses, its surface brightness
has dropped, making it less prominent in the most recent images. 

\begin{table*}[t]
\centering
\caption{Multi-roll {\it HST} coronagraphic observations of the Fomalhaut system\label{tab:data}}
\begin{tabular}{lllllrrr}
Program ID & Date & Instrument & Filters & Apertures & $N_{\rm Images}$ & $N_{\rm Rotations}$ & PSF \\
\midrule
GO9862$\checkmark$  & 2004-May|Aug    & ACS  & F814W             & CORON1.8                 & 5  & 2           & Vega\\
GO10390$\checkmark$ & 2004-Oct       & ACS  & F606W,F814W       & CORON1.8                 & 32 & 3           & Vega\\
GO10598$\checkmark$ & 2006-Jul       & ACS  & F435W,F606W,F814W & CORON1.8,CORON3.0        & 53 & 4           & Vega\\
GO11818$\checkmark$ & 2010-Jun|Sep    & STIS & N/A               & WEDGEB2.5$^{\ast}$       & 19 & 7$\dagger$  & ADI \\
GO12576$\checkmark$ & 2012-May      & STIS & N/A               & WEDGEB2.5                & 48 & 12$\dagger$ & ADI \\
GO13037             & 2013-May          & STIS & N/A               & WEDGEB2.5                & 48 & 12$\dagger$ & ADI\\
GO13726             & 2014-Sep          & STIS & N/A               & WEDGEB2.5$^{\ast}$       & 24 & 8           & Vega \& ADI\\
\bottomrule
\end{tabular}
\begin{flushleft}
\addtabletext{Programs marked with ``ADI'' were planned to use angular differential imaging and therefore
did not observe a PSF source.}

\addtabletext{$\dagger$The majority of rotations were separated by only a few degrees ($\le 25^{\circ}$).}

\addtabletext{$\checkmark$Previously published data.}

\addtabletext{$^{\ast}$Observations were also taken at additional apertures (BAR10 for GO11818 and BAR5 for GO13726), but were
not included due to low S/N at the position of Fomalhaut b.}
\end{flushleft}
\end{table*}

\section*{Archival Data}

Fomalhaut b has been observed only in scattered light and only with {\it HST}. The stable 
optical system of {\it HST} and the lack of atmospheric disturbance enable high-contrast imaging at visible wavelengths with 
the aid of coronagraphs. Fomalhaut b was discovered in images taken with the Advanced Camera for Surveys (ACS) coronagraph 
in 2004 and 2006 \citep{kalas08}. The failure of the High Resolution Channel (HRC, which included the coronagraph) of ACS in 2007
led to the Space Telescope Imaging Spectrograph (STIS) becoming the single coronagraphic instrument onboard {\it HST}; 
thereafter the monitoring of Fomalhaut b continued with STIS. Images of Fomalhaut b taken with STIS in 2010 and in 2012 
have been published by \cite{kalas13}. 
 
The coronagraphs in these instruments differ in design and performance. The ACS Lyot coronagraph was located in the aberrated 
beam of the telescope. This limited the performance of the coronagraph at smaller ($\le 3^{\prime\prime}$) inner working angles. 
However, it was able to observe using a suite of optical filters. The STIS coronagraph is located in the 
corrected beam; however, it is unfiltered and therefore detects photons from 0.4 to 1 $\mu$m. While this yields high sensitivity, 
it also limits the fidelity of the chromatically dependent point spread function (PSF) of the telescope, which must be observed 
close in time to the target star because of temporal drifts in the image.  

Since 1999, 11 {\it HST} coronagraphic programs observed Fomalhaut, either with ACS or with STIS. Of these, only 7 were 
true high-contrast imaging, multiorbit, and multirotation-angle programs as described in Table \ref{tab:data}; the remaining 
4 programs took only a single or a few images. Here, we describe the previously unpublished 2013 and 2014 observations of 
Fomalhaut b and provide a coherent and independent rereduction and analysis of all the useful {\it HST} data.

\subsection*{Rereduction of Archival ACS Data}

Three of the 11 {\it HST} coronagraphic programs used ACS (PI: Paul G.\ Kalas) and provided the discovery images
of Fomalhaut b \citep{kalas08}. We first give a brief description of the observations and the observing techniques employed 
within these observations. We then describe our rereduction of these data, carried out so all of the available photometry is treated identically.

Each ACS program observed Fomalhaut and Vega at multiple roll angles, the latter star to define the PSF. While Vega and Fomalhaut 
are not exact color matches, the response of the system through color filters to each source was approximately the same. Various 
length exposures were taken of each source to enable imaging of the inner and outer regions without saturation and to high 
signal-to-noise ratios. 

The publicly available data for these observations were obtained from the Space Telescope Science Institute (STScI) Mikulski 
Archive  for Space Telescopes (MAST) website. We downloaded corrected ($\tt{\_drz.fits}$) files, which rectify the non-square 
pixels of the HRC detector and also correct the field distortions. In our reductions, we rejected observations with issues 
such as poor centering on the coronagraph and also combined measurements in the same filter taken sufficiently closely in time that 
the relative motion of Fomalhaut b is insignificant (GO9862 and GO11818). 

Over an orbit, the Hubble Space Telescope undergoes thermal expansions (a.k.a.\ ``breathing''), which vary the PSF. Additionally, the ACS
coronagraph transmits a small fraction of the occulted flux; the transmitted PSF  depends on the total count received. Therefore, the
ACS observations were target-PSF paired based on exposure length and
inter-orbit sequence. PSF subtraction residuals can become substantial if target observations are not registered to $\le 0.1$ px with 
their PSFs.  We determined the shifts between each image and its chosen PSF by observing the subtraction residuals by eye. Using the 
software IDP3, we registered the images well within 0.1 px by minimizing the radial streak pattern over a radius of  $10^{\prime\prime}$ 
from the occulter. We also registered the shifts between the target images in a similar manner. We found an average offset in $x$ and $y$ 
pixel coordinates of $\Delta x = -0.32 \pm 0.21$ and $\Delta y = 0.35 \pm 0.42$, which is close to the nominal quarter pixel value 
given in the Instrument Handbook for pointing precision. The fluxes of the PSFs were scaled for each pairing by determining the 
background levels at the edges of the fields. 

To pinpoint the location of Fomalhaut, which then defines the derotation center of the images, we determined the rotational symmetry 
origin of each image. To do so, we rotated each image by 180$^{\circ}$ and subtracted it from its original version. Subtraction residuals 
within the occulting spot became apparent at $\sim \pm 1$ px offsets. We use this value, 
along with the rotational angles of individual images and the full-width at half-maximum (FWHM) of the target PSFs, to estimate astrometric 
errors. Following PSF subtraction, subtraction masking, and derotation, the individual target images were combined.

\begin{figure*}[!ht]
\centering
\includegraphics[width=.99\linewidth]{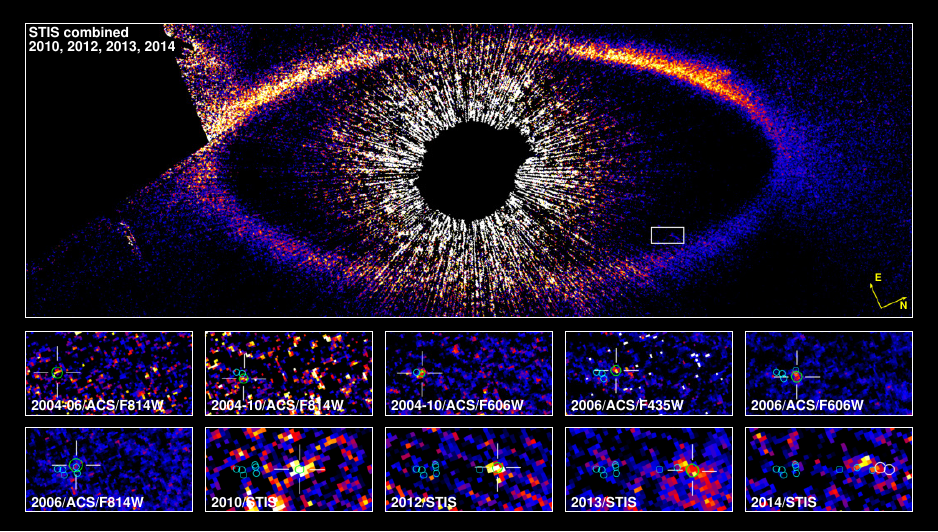}
\caption{({\it Top}) Median combined image of all multi-roll STIS observations (2010, 2012, 2013, and 2014) of the Fomalhaut 
system. Fomalhaut b is visible in the image within the white rectangle, which is $2^{\prime\prime}\times1^{\prime\prime}$ in 
size and highlights the area shown in the postage-stamp images below. The postage-stamp images show the individual 
observations (see text for descriptions of each). The individual images are scaled to the same level per filter 
[0 to 0.1 counts per second (cts$\cdot$s$^{-1}$) for the F606W and F435W ACS filters and for STIS and 0 to 0.05 cts$\cdot$s$^{-1}$ 
for the F814W ACS filter]. The green circles with crosses highlight the then
current positions of Fomalhaut b, with 3$\sigma$ astrometric error radii, while the smaller cyan color circles show the previous positions,
to highlight the spatial motion of the source. For the 2014 image, we show the two locations predicted by the two independent trajectory fits.
The bright spot ``near'' the predicted locations is too far to be considered associated with Fomalhaut b.}
\label{fig:Bigplot}
\end{figure*}

\subsection*{(Re)Reduction of Archival and Additional STIS data}

Four programs using STIS employed a multiroll/multiorbit observation technique (PI: P.\ Kalas for all of them), which is necessary to achieve 
high-contrast imaging. 
The results from the first two  programs have been published \citep{kalas13}; here we publish the results of the last two. As 
STIS is unfiltered for coronagraphic imaging, precise color-matching of the PSF source is critical. Since well-matched PSF calibrators 
may not be available nearby, it is common to employ the angular differential imaging (ADI) technique, where a star becomes its own PSF 
calibrator by combining images of it taken at different roll angles. This approach works well for edge-on or inclined narrow-belt disk systems, 
such as that of Fomalhaut. 

The STIS coronagraph consists of two occulting wedges, and two occulting bars. The GO11818 program observed Fomalhaut in 2010 using the 
2.5$^{\prime\prime}$ wide position on the ``B'' wedge. Deeper images at the same occulting position were obtained in 2012 (GO12576) and 
2013 (GO13037) utilizing additional roll angles (for a total of 12), which helps to suppress PSF subtraction residuals as well as 
establishing a better PSF for ADI. The 2014 observations (GO13726) integrated for a slightly shorter time, but the STIS observations 
are contrast and not photon limited, so the data are still adequate for the detection of Fomalhaut b. 

The reduction of the STIS data followed similar steps to those for the ACS data. The level of geometric distortion for
STIS is minimal; the largest offset near the location of Fomalhaut b is at most 0.5 px, based on available 
distortion maps \citep{walsh01}. For high-quality astrometric measurements, however, subpixel alignment precision is necessary.
Therefore, we opted to download and analyze the distortion corrected images ($\tt{\_sx2.fits}$). 
Each of the images was used to generate its own mask, based on the underlying STIS occulter mask and generous saturation 
masking at 75\% of the full well value. We used the same methods as for the ACS data to measure the offsets among all images within an 
observation program. The tracking of radial subtraction patterns and their minimization by eye allowed us to define shifts in 
both pixel coordinates within 0.1 px. As the STIS occulters are not transparent, its coronagraphic PSF is more stable than that of ACS, 
and the radial patterns remain generally the same for the complete observing program. This enables high-precision registering of all 
data to the same coordinates. We did, however, notice small interorbit variations and therefore decided to construct PSFs for each 
exposure sequence ID, median combining all first, second, etc.\ images of all orbits. This method follows the ADI technique,
but also folds in the thermal changes to the HST optical telescope assembly (OTA).

The Lyot stop for the STIS coronagraph blocks only light from the edges of the OTA; therefore, the diffraction spikes are prominent 
in the images and only the wings of the PSF are suppressed. While this necessitates additional masking during image processing and 
reduces the imaging real estate, it also allows precise locating of the central star for image derotation. 
Following image registering with PSF subtraction residuals, the location of the star (determined by tracking the diffraction
spikes), has a remarkably small error of 0.12 px (6 mas) in both coordinates. As for the ACS images, following PSF subtractions, 
masking, and derotation, the individual images were combined.

In Fig.\ \ref{fig:Bigplot}, we show the results of our image reductions and analysis, including a high signal-to-noise
combined STIS image and detailed snapshots of Fomalhaut b at all epochs.

\section*{Astrometry and Photometry of Fomalhaut b}

\begin{table}[t]
\centering
\caption{Astrometry of the Fomalhaut b object\label{tab:astr}}
\begin{tabular}{llrr}
Date & Instrument &  $\Delta$R.A.\ ($^{\prime\prime}$) & $\Delta\delta$ ($^{\prime\prime}$) \\
\midrule
2004/Jun & ACS (F814W)   & -8.542 $\pm$ 0.021 & 9.144  $\pm$ 0.021 \\
2004/Oct & ACS (F606W)   & -8.580 $\pm$ 0.011 & 9.198  $\pm$ 0.011 \\
2004/Oct & ACS (F814W)   & -8.642 $\pm$ 0.017 & 9.194  $\pm$ 0.018 \\
2006/Jul & ACS (F435W)   & -8.614 $\pm$ 0.020 & 9.363  $\pm$ 0.020 \\
2006/Jul & ACS (F606W)   & -8.683 $\pm$ 0.021 & 9.341  $\pm$ 0.021 \\
2006/Jul & ACS (F814W)   & -8.590 $\pm$ 0.025 & 9.364  $\pm$ 0.026 \\
2010/Sep & STIS          & -8.850 $\pm$ 0.016 & 9.824  $\pm$ 0.016 \\
2012/May & STIS          & -8.915 $\pm$ 0.019 & 10.024 $\pm$ 0.020 \\
2013/May & STIS          & -9.018 $\pm$ 0.027 & 10.173 $\pm$ 0.025 \\
\hline
2014/Sep & (Bound model) & -9.029$\dagger$    & 10.273$\dagger$ \\
2014/Sep & (Cloud model) & -9.093$\dagger$    & 10.360$\dagger$ \\
\bottomrule
\end{tabular}
\begin{flushleft}
\addtabletext{$\dagger$ Projected location of the source (no detection).}
\end{flushleft}
\end{table}

\begin{figure}[!ht]
\centering
\includegraphics[width=.99\linewidth]{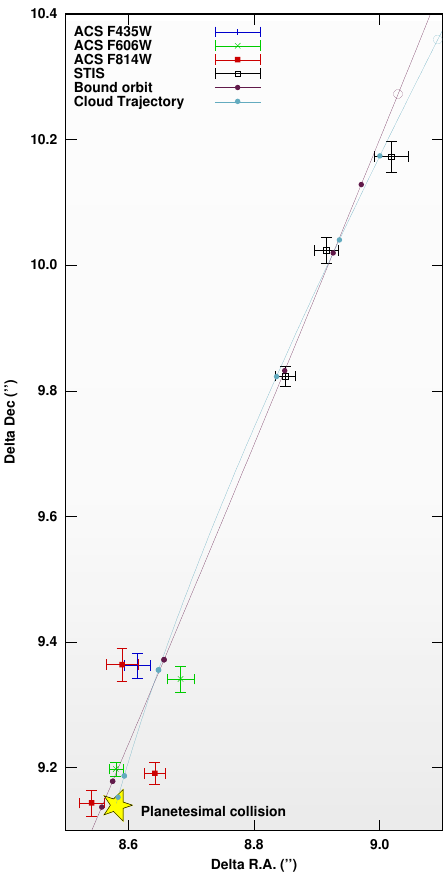}
\caption{Astrometry of Fomalhaut b (see text for details). The circles show the 
projected location of the object in 2014 for both the bound orbit and cloud models. The purple
line and dots give the best-fitting bound orbit and locations for each observation, while 
the cyan line and dots give the same for the dust cloud model.}
\label{fig:astrometry}
\end{figure}

Determining the astrometric location of Fomalhaut b in the ACS and STIS images started from the published coordinates 
\citep{kalas13} in the 2004, 2006, 2010, and 2012 images and was based on forward projections for the 2013 and 2014 images. 
While the source is clearly identifiable in the F606W ACS and the 2010 and 2012 STIS images, it is more difficult to pinpoint it
in the others. For the F435W and F814W images, we searched for peaks near the F606W positions, while for the 2013 STIS image
forward projections from the previous epochs aided in locating the source. We calculated the locations using the centroiding 
algorithm in IRAF,  DAOPHOT.APPHOT, initiated on the brightest pixel within a 2 pixel radius of the predicted location. 
The object is visibly expanding in size in the STIS dataset and therefore its astrometric position has an increasingly 
larger error. The object was not detected in the 2014 dataset at either model trajectory locations. All peaks visible in 
the 2014 dataset are spatially far enough from any realistic trajectory projection that they can be dismissed. 
The astrometric positions are summarized in Table \ref{tab:astr} and plotted in Fig.\ \ref{fig:astrometry}. 

The astrometric errors are calculated as the root sum squares of the derotational errors and the PSF FWHM astrometric errors.
The derotational errors in right ascension and declination were calculated for each individual observation, based on the 
derotation angles within each observation set and the precision of image centering, weighted by the exposure time of 
each individual image. We calculated the derotational errors with a Monte Carlo code. The centroiding astrometric errors were calculated
based on the standard formula of
\begin{equation}
    \sigma_{\rm centroid} = \frac{1}{2.355}\frac{\rm FWHM}{\rm SNR}\;,
\end{equation}
where SNR gives the peak pixel signal to background pixel noise ratio. Centroiding FWHMs were provided by IRAF for both right
ascension and declination. 

We performed aperture photometry of Fomalhaut b in all of the final reduced images. The photometry was performed with a 2.1 pixel 
radius aperture. The background was measured in 2.1 pixel radius apertures randomly 
placed at the same stellocentric distance as the target source (avoiding areas where the disk is prominent). The photometry 
error was determined as the standard deviation of these background measurements. This is the same method that \cite{kalas13} 
used to estimate the background and photometry errors. We applied aperture corrections of factors of 3.03, 2.56, 2.27, and 1.45 
for the ACS (F814W, F606W, F435W) and STIS observations, respectively, based on TinyTim theoretical PSFs 
\citep{krist11}, which include broadening due to the Lyot stops.

\begin{table*}[t]
\centering
\caption{Photometry of the Fomalhaut b object\label{tab:phot}}
\begin{tabular}{llllll|l}
Date & Filter   & F (cts s$^{-1}$)           & F (Jy) & m$_{\rm Vega}$ & m$_{\rm AB}$ & F$_{\rm prev}$ \\
\midrule
2004/Jun & F814W & 0.65 $\pm$ 0.16            & 3.83e-7 & 24.51 & 24.94 & N/A                                              \\
2004/Oct & F606W & 1.70 $\pm$ 0.22            & 5.23e-7 & 24.51 & 24.61 & 24.43 $\pm$ 0.08 mag (1); 24.92 $\pm$ 0.10 mag (2); 6.3 $\pm$ 1.0 $\times 10^{-7}$ Jy (3)\\
2004/Oct & F814W & 0.72 $\pm$ 0.23            & 4.23e-7 & 24.40 & 24.83 & N/A                                          \\
2006/Jul & F435W & 1.28 $\pm$ 0.26            & 8.90e-7 & 24.11 & 24.03 & 25.22 $\pm$ 0.18 (2); 3.6 $\pm$ 0.9 $\times 10^{-7}$ Jy (3) \\
2006/Jul & F606W & 1.52 $\pm$ 0.15            & 4.71e-7 & 24.63 & 24.72 & 25.13 $\pm$ 0.09 mag (1); 24.97 $\pm$ 0.09 mag (2); 4.3 $\pm$ 0.6 $\times 10^{-7}$ Jy (3) \\
2006/Jul & F814W & 0.56 $\pm$ 0.08            & 3.24e-7 & 24.69 & 25.12 & 24.55 $\pm$ 0.13 mag (1)  24.91 $\pm$ 0.20 mag (2); 3.6 $\pm$ 0.7 $\times 10^{-7}$ Jy (3) \\
2010/Sep & Clear & 0.77 $\pm$ 0.10            & 3.58e-7 & 24.77 & 25.02 & 0.49 $\pm$ 0.14 cts s$^{-1}$ (4); 6.1 $\pm$ 2.1 $\times 10^{-7}$ Jy (3)\\
2012/May & Clear & 0.58 $\pm$ 0.10            & 2.71e-7 & 25.07 & 25.32 & 0.50 $\pm$ 0.11 cts s$^{-1}$ (4)\\
2013/May & Clear & 0.47 $\pm$ 0.08            & 2.17e-7 & 25.32 & 25.56 & N/A                         \\
\hline
2014/Sep$\dagger$  & Clear & 0.32 $\pm$ 0.14  & 1.48e-7 & 25.73 & 25.97 & N/A                         \\
2014/Sep$\ddagger$ & Clear & 0.23 $\pm$ 0.14  & 1.07e-7 & 26.09 & 26.33 & N/A                         \\
\bottomrule
\end{tabular}
\begin{flushleft}
\addtabletext{References: 1) \cite{kalas08}, 2) \cite{currie12}, 3) \cite{galicher13}, 4) \cite{kalas13}}\\
\addtabletext{$\dagger$Flux at the predicted location of the source using the bound orbit trajectory.}\\
\addtabletext{$\ddagger$Flux at predicted location of the source using the unbound cloud model trajectory.}
\end{flushleft}
\end{table*}

We summarize our photometry compared to previously published values in Table \ref{tab:phot}. The instrumental 
flux values (counts$\cdot$s$^{-1}$) were converted to various physical flux units using {\tt synphot}, assuming the incident light that is scattered has a spectral 
distribution identical to that of the host star (A3V) and using conversion factors appropriate for the instrument setup and 
observation date. The pivot wavelength of the instrument/filter setup was used to calculate the Jansky (Jy) units. Our photometry values 
are generally in agreement with the previously published values for the ACS data, except for the single observation in the F435W filter. 
For STIS, our photometry agrees with that of \citep{kalas13}; however, it is fainter than that of \citep{galicher13} by a factor of 2. Such 
discrepancies have been noted in the literature for this source and are likely an outcome of differences in data reduction and 
photometry methods. Nevertheless, the general conclusions of our paper are not affected by these discrepancies because we have applied 
identical procedures to all the images.

\begin{figure}[!t]
\centering
\includegraphics[width=.99\linewidth]{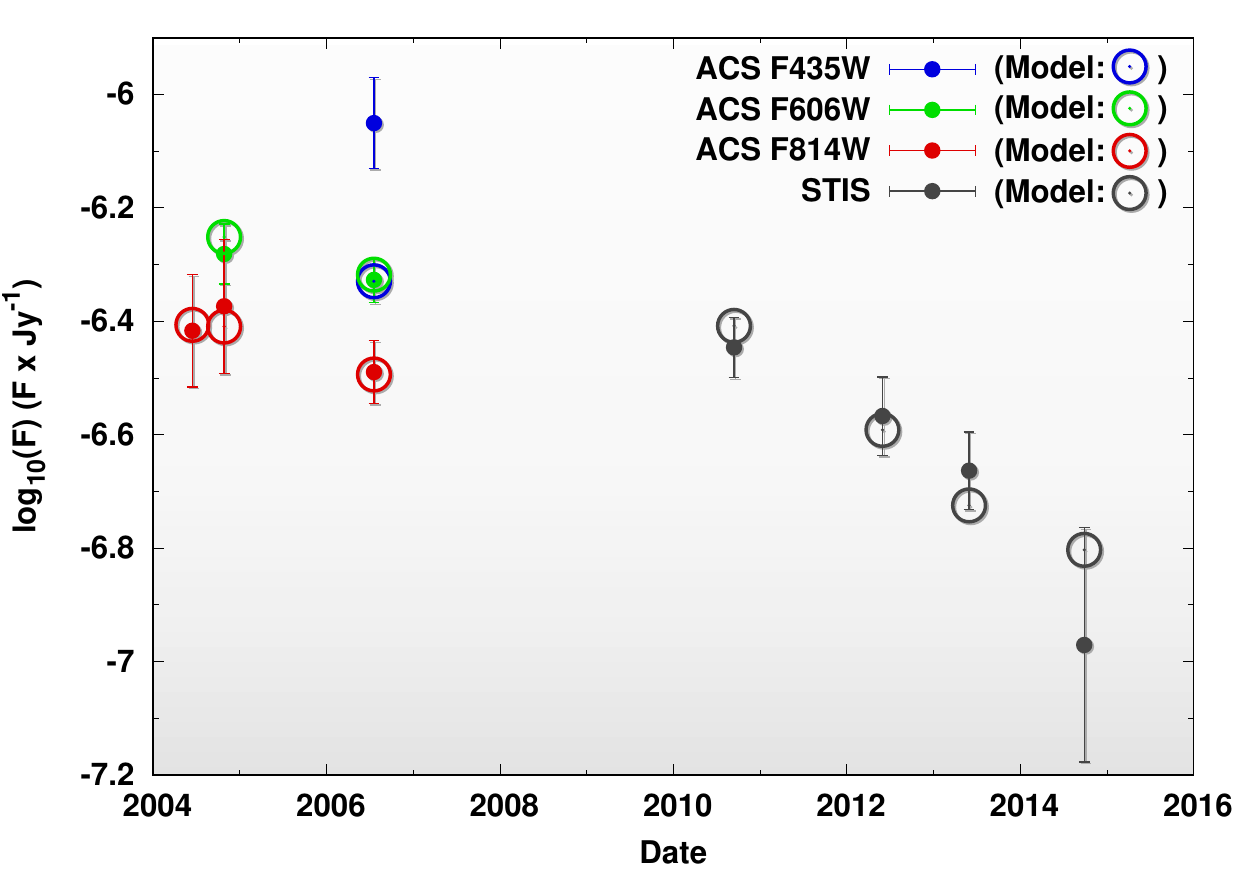}
\caption{The photometry of Fomalhaut b at all observed epochs and wavelengths. The error bars shown are at 
1$\sigma$. The offset between the ACS and STIS data is most likely an artifact, due to the larger STIS pixelsizes, 
the spatial expansion of the dust cloud, and the method of photometry that was necessary. The expansion of the dust 
cloud over time, convolved with the HST PSF, results in the central peak brightening. This artifact is further 
enhanced by the larger STIS pixel size. As we had to use larger physical apertures for the STIS data than for the 
ACS data (same number of pixels), the object seems to brighten artifically. Identical photometry of the cloud 
model ACS and STIS data, shown with empty circles, exhibits similar behavior. }
\label{fig:photometry}
\end{figure}

The object fades in each individual band over time; however, there is a color shift between the various ACS filters and an offset
between the ACS and STIS data in general. The color shift is a natural consequence of the dust cloud being slightly bluer than the 
central star. The offset between the ACS and STIS data is most likely an artifact, brought on by the combination of changes in two 
variables: 1) the source becomes extended over time and 2) the STIS photometry aperture is physically larger than the ACS one, due 
to the larger projected pixel sizes of the detector (Fig.\ \ref{fig:photometry} legend). The precision alignment of the STIS data and lack of any higher 
signal points near the predicted location of the source (even with a wide margin of astrometric error) gives high confidence to 
our non-detection in 2014.

\section*{Interpretations of the Observations}

We draw three basic conclusions about Fomalhaut b from the observations: 1) It is probably 
moving out of the system, 2) it has become increasingly extended, and 3) it has faded below our detection limits. 
We support these conclusions first by comparing bound and unbound fits to the observed motion of the object. We then analyze 
its evolution in size and brightness using a model providing a self-consistent explanation of all three unique aspects of its behavior.

\subsection*{Bound Orbit of Fomalhaut b}

\begin{figure}[!t]
\centering
\includegraphics[width=0.99\linewidth]{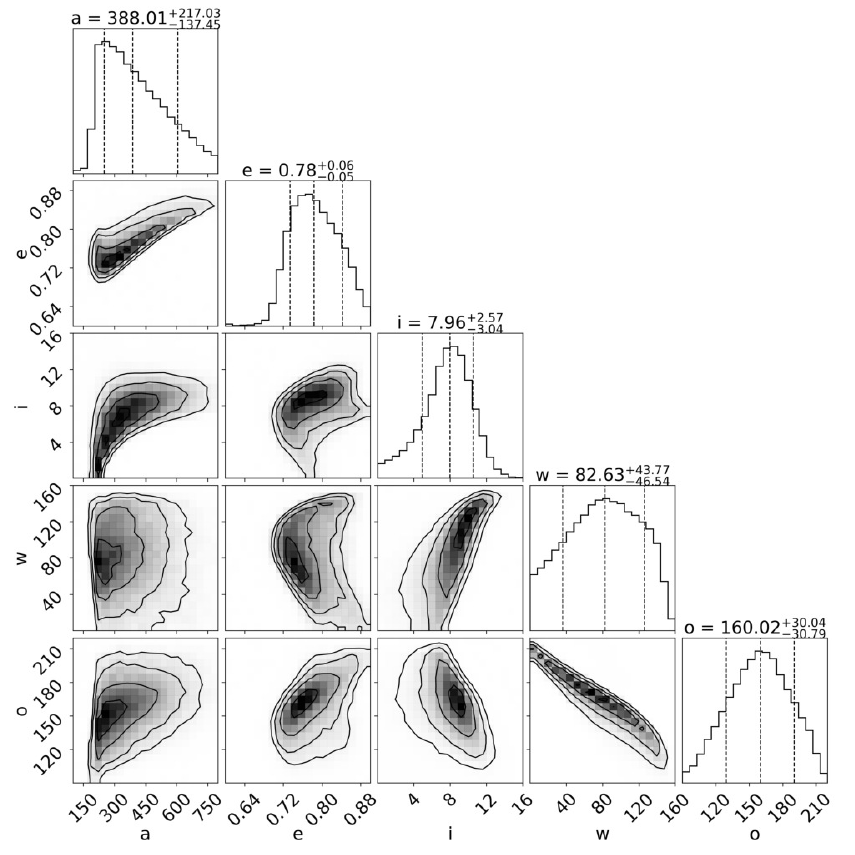}
\includegraphics[width=0.99\linewidth]{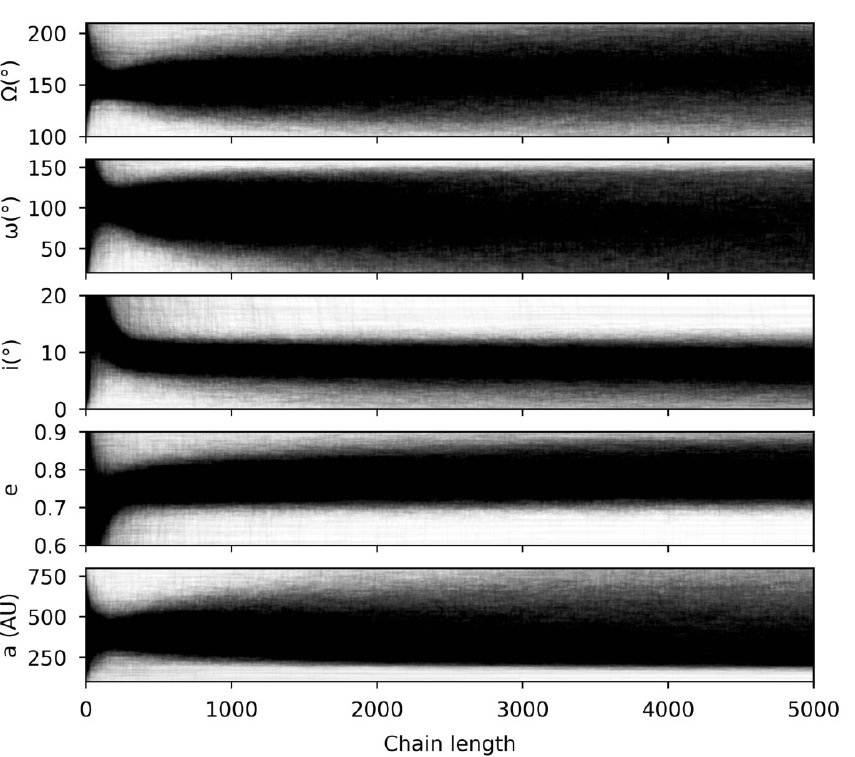}
\caption{{\it Top:} Posterior probability distributions over the free orbital parameters of Fomalhaut b, 
assuming a stable bound orbit, using MCMC analysis (6000 chains, 5000 steps, burn-in limit at 2000 steps). The diagonal panels show 1-D 
projections (marginalized over all other parameters) of the probability density, while the off-diagonals 
show 2-D projections of the correlations between parameters. The mean and the 1$\sigma$ levels of the 1-D 
projections are given in the figure, while the global reduced $\chi^2 = 3.18$ minimum of the distribution is at
$a = 385.5~{\rm au}$, $e = 0.7604$, $\iota = 9.90^{\circ}$, $w = 118.00^{\circ}$, $\Omega = 139.59^{\circ}$.
{\it Bottom:} The MCMC chains of the fitted variables and their convergence for the bound orbit model.}
\label{fig:corner}
\end{figure}

If the underlying object for Fomalhaut b is planetary, it should be following a stable bound orbit. To fit the the best solution for such 
an orbit and its joined Bayesian errors, we used a Markov Chain Monte Carlo (MCMC) fitting 
routine \citep{foreman13}. Orbital solutions were projected forward to each epoch from the orbital point closest to the first 
observation location. We assumed uniform priors on the orbital elements (semi-major axis $a$, eccentricity $e$, inclination 
$\iota$, argument of periapsis $w$,
right ascension of the ascending node $\Omega$) and determined their statistics using 5000 steps following a burn-in limit 
of 2000 steps for 6000 test chains. These orbital elements are defined in the plane of the Fomalhaut disk (assuming a 
system position angle of $336^{\circ}$ and inclination of $66^{\circ}$). In Fig.\ \ref{fig:corner}, we show the results.

In Fig.\ \ref{fig:astrometry}, we also plot (with purple color) the best orbital solution (at the reduced $\chi^2$ minimum, given in the
Fig.\ \ref{fig:corner} legend) assuming this stable bound orbit. Although the observations lie near the  track of the ``best'' 
orbit, the positions along that track do not agree with the observations within errors. This issue is prominent for 2013, the point 
to the upper right, where the observed position is significantly ahead of the bound orbit prediction (purple dot, Fig.\ \ref{fig:astrometry}), consistent with the 
object experiencing nongravitational acceleration directed away from the star. A planetary body would also not explain the fading 
and extent of the image. Therefore, additional models of the motion and behavior of the source are necessary.

\subsection*{Fomalhaut b as a Dynamically Dissipating Collisional Dust Cloud}

We now consider an alternative model tracing the evolution of the system through the following steps: 1) Two large ($\ge$ 100 
km in radius) asteroids collide catastrophically; 2) the collision fragment sizes follow a power-law distribution down to the submicrometer level;
3) the fragments inherit random velocities, resulting in the expansion of the dust cloud relative to the center of mass (COM)
of the colliding asteroids; and 4) the final trajectories depend on fragment size: larger ones unaffected by radiative
forces will undergo Keplerian shear, while the smaller ones will converge onto radial trajectories and leave the system. 
The surface area of the dust cloud observed in scattered light is dominated by
the submicrometer particles that are leaving the system. The apparent motion of the observed image should reflect such motion.

To test this scenario, we modeled the dynamical evolution of such a hypothetical system using our code 
$\tt DiskDyn$. Developed to model the dynamical evolution of debris disks and dust
clumps, {\tt DiskDyn} is a complex numerical code able to include the effects of gravitational, radiative, and magnetic 
forces on dust particles, as well as calculate the gravitational effects of a large number of massive bodies. Additionally,
{\tt DiskDyn} provides {\tt .fits} images of the scattered light and thermal emission from the dust particles and 
complete SEDs. Due to the numerically intensive calculations, {\tt DiskDyn} runs on Graphical Processing Units (GPUs).

\begin{figure}[!t]
\centering
\includegraphics[width=.99\linewidth]{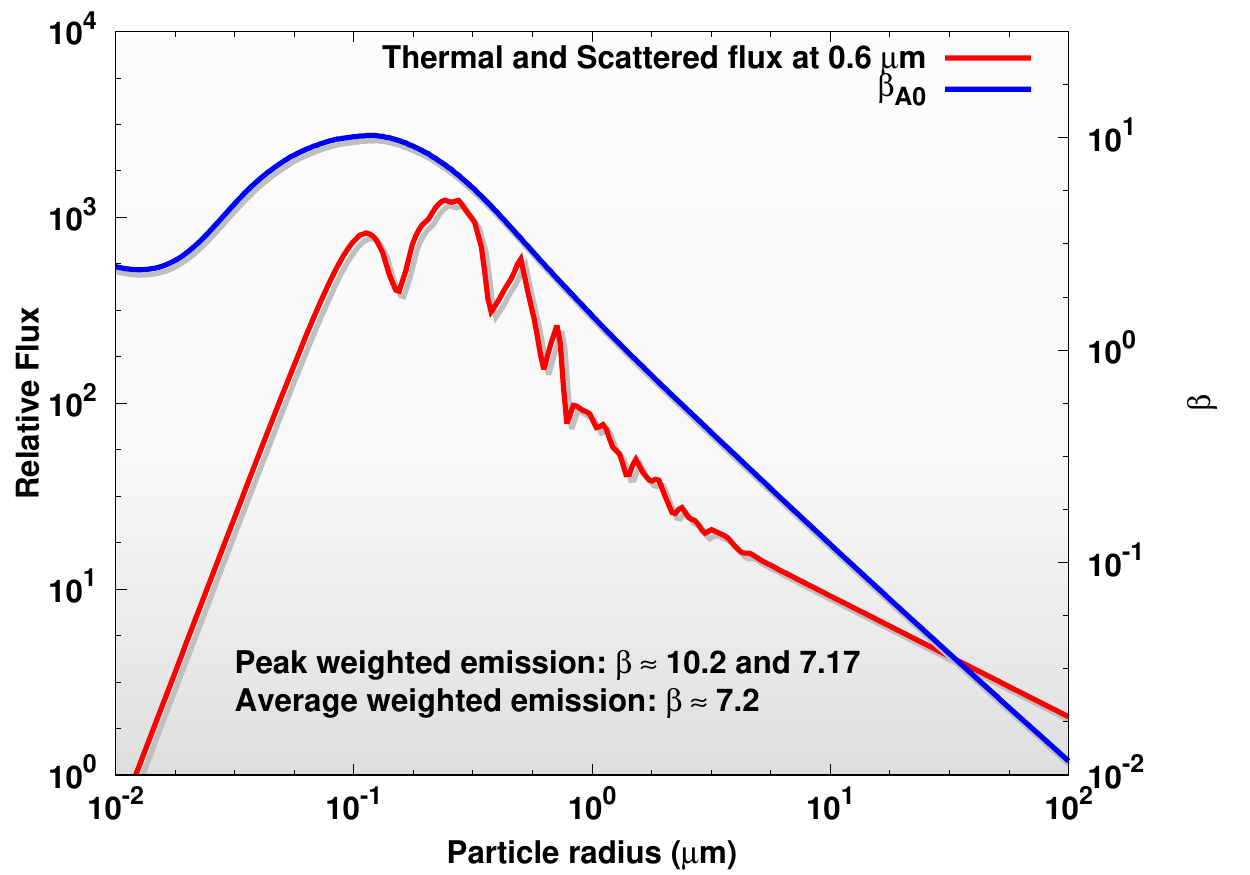}
\caption{The relative emission flux (thermal and scattered) at an observational wavelength of 0.6 $\mu$m 
(approximating the F606W ACS filter and the peak of the spectral response of the unfiltered STIS detector) 
as a function of particle size, assuming the modeled underlying power-law size distribution. We also plot the 
value of $\beta$ as a function of particle size on the secondary axis for the dust particles we modeled around
an A0 spectral-type star. The particles contributing most of the flux have a weighted average of $\beta=7.2$.
}
\label{fig:beta}
\end{figure}

While $\tt DiskDyn$ is a versatile modeling code, it does not provide fitting tools. Therefore, we first fitted the 
trajectory of the dust cloud 
in a similar fashion as for the bound orbit solution (with MCMC) to obtain the input values for $\tt DiskDyn$; however, 
for this fit, we did not constrain the trajectory to be bound. Instead, the launch location and velocity of the cloud were 
defined by its Cartesian coordinates and velocities (yielding an extra free parameter relative to the bound orbit solution). 
The trajectories of small dust grains are characterized by $\beta$, the ratio of radiation force to gravitational force on 
the particles. In Fig.\ \ref{fig:beta}, we show the total (thermal and scattered) flux as a function of particle size 
(weighted by the dust number density) at an observational wavelength of 0.6 $\mu$m, which corresponds to the wavelength 
observed with ACS in the F606W filter and with STIS. Fig.\ \ref{fig:beta} also shows the $\beta$ values, as a function of particle 
radius. The particle radius range most prominently observed with {\it HST} is between 0.07 and 0.7 $\mu$m, with flux peaks 
from dust with radii of 0.11 and 0.23 $\mu$m. These sizes have $\beta$ values of 10.2 and 7.17, respectively. Therefore, 
for initial trajectory fitting we assumed $\beta=10$ for all particles. However, our final $\tt DiskDyn$ dust cloud model 
calculates realistic dust optical properties [since we are only concerned with scattered light approximated by Mie theory, 
we used astronomical silicates only \citep{ball2016}] and dynamics.

\begin{figure}[!t]
\centering
\includegraphics[width=0.99\linewidth]{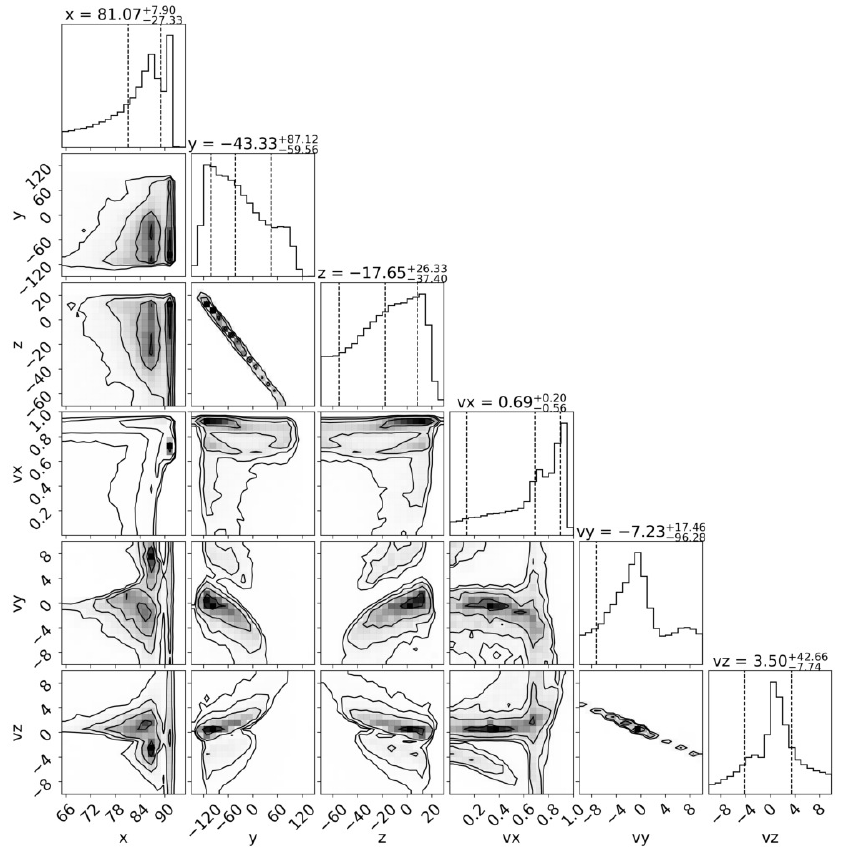}
\includegraphics[width=0.99\linewidth]{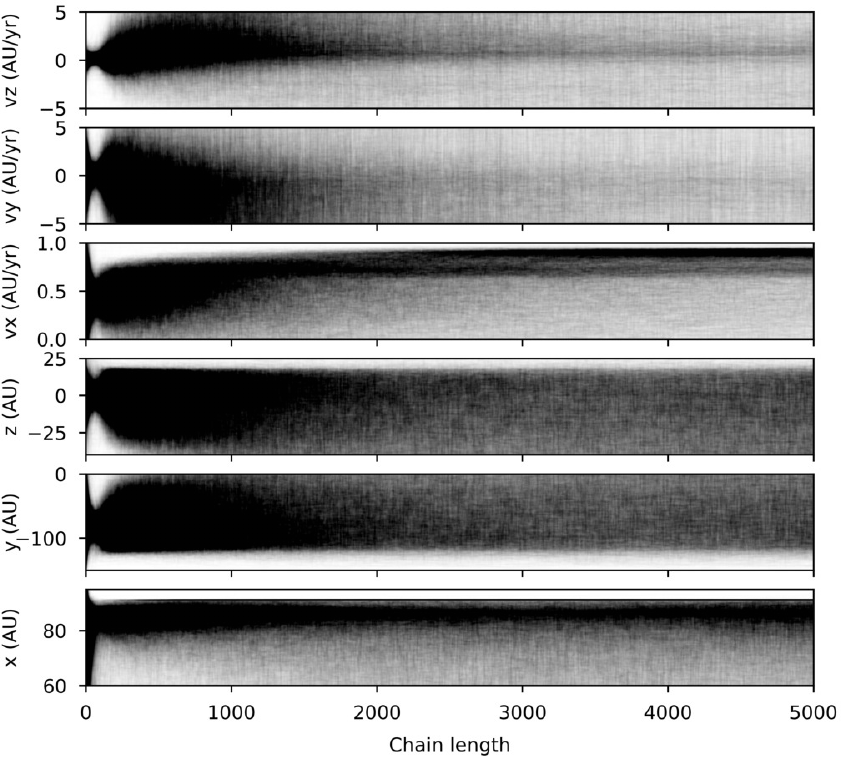}
\caption{{\it Top}: Posterior probability distributions over the free orbital parameters of Fomalhaut b, 
assuming an unconstrained cloud model orbit, using MCMC analysis (6000 chains, 5000 steps, burn-in limit at 2000 steps). 
The global reduced minimum of the distribution is $\chi^2 = 2.38$. The coordinates (in au) and velocities (in au yr$^{-1}$) 
are defined in the plane of the Fomalhaut system. {\it Bottom}: The MCMC chains of the fitted variables and their convergence for the unconstrained cloud model.}
\label{fig:corner2}
\end{figure}

Fig.\ \ref{fig:corner2} shows the MCMC analysis of the initial location 
of the planetestimal collision and launch velocity of the observed dust particles, determined in the plane of the Fomalhaut system. 
The only constraint on the initial condition was that the collision had to  occur within 150 au of Fomalhaut. 
As Fig.\ \ref{fig:corner2} shows, two families of solutions were found, one with a sharp probability peak at 
${\rm x}=91~{\rm au}$, and one with a broader distribution ($80\le {\rm x} \le 90~{\rm au}$). The best fit (with a reduced $\chi^2=2.38$ 
- compared with 3.18 for the best bound orbit fit) was for a collision occuring at ${\rm x} = 91.24~{\rm au}$, ${\rm y} = -47.62~{\rm au}$, 
and ${\rm z} = -13.47~{\rm au}$, with an initial 
launch velocity of 1.59 au$\cdot$yr$^{-1}$, twice the Keplerian orbital velocity for a circular orbit at the location of the 
collision.  However, solutions with velocities between 0.59 and 22.13 au$\cdot$yr$^{-1}$ (at different launch positions) produce fits with 
reduced $\chi^2$ lower than the best bound orbit fit. The best fitting trajectories were from the first family of solutions, typically 
having larger velocities (from 1.08 au$\cdot$yr$^{-1}$), while the best fits from the second family of solutions yielded fits with reduced 
$\chi^2$ similar to that of the bound orbit with velocities up to 0.8 au$\cdot$yr$^{-1}$ (i.e.\ sub-Keplerian velocities, assuming circular orbits). 
Trajectories originating from elliptical bound orbits were found in both families.

\begin{figure*}[!t]
\centering
\includegraphics[width=0.73\textwidth]{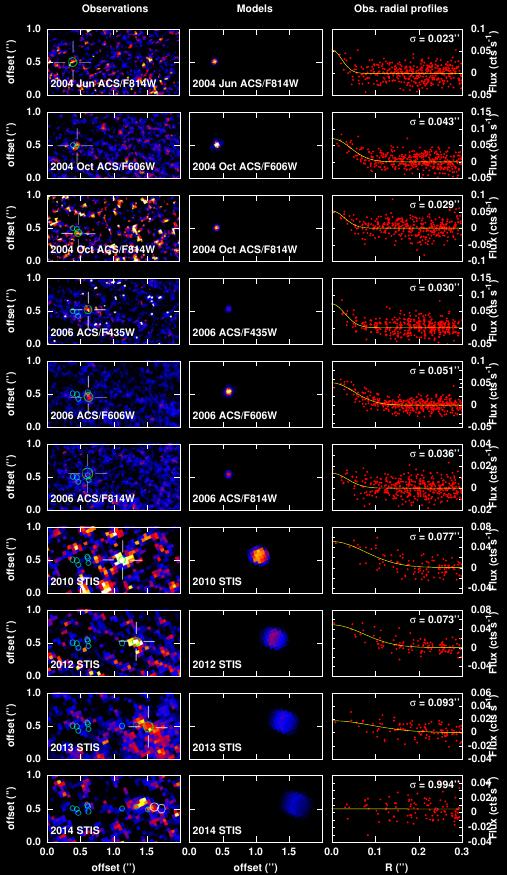}
\caption{Comparison of the evolution of the spatial scale, location, and surface brightness of the dust cloud as observed and modeled. 
The scaling of the images are the same as in Fig.\ \ref{fig:Bigplot} for both models ({\it Center} column) and observations ({\it Left} column).
{\it Right} column also shows the fitted radial profiles of the observations. The radial profiles are corrected for instrumental effects in Fig.\ \ref{fig:sigma}.}
\label{fig:radial}
\end{figure*}

\begin{figure}[!t]
\centering
\includegraphics[width=0.99\linewidth]{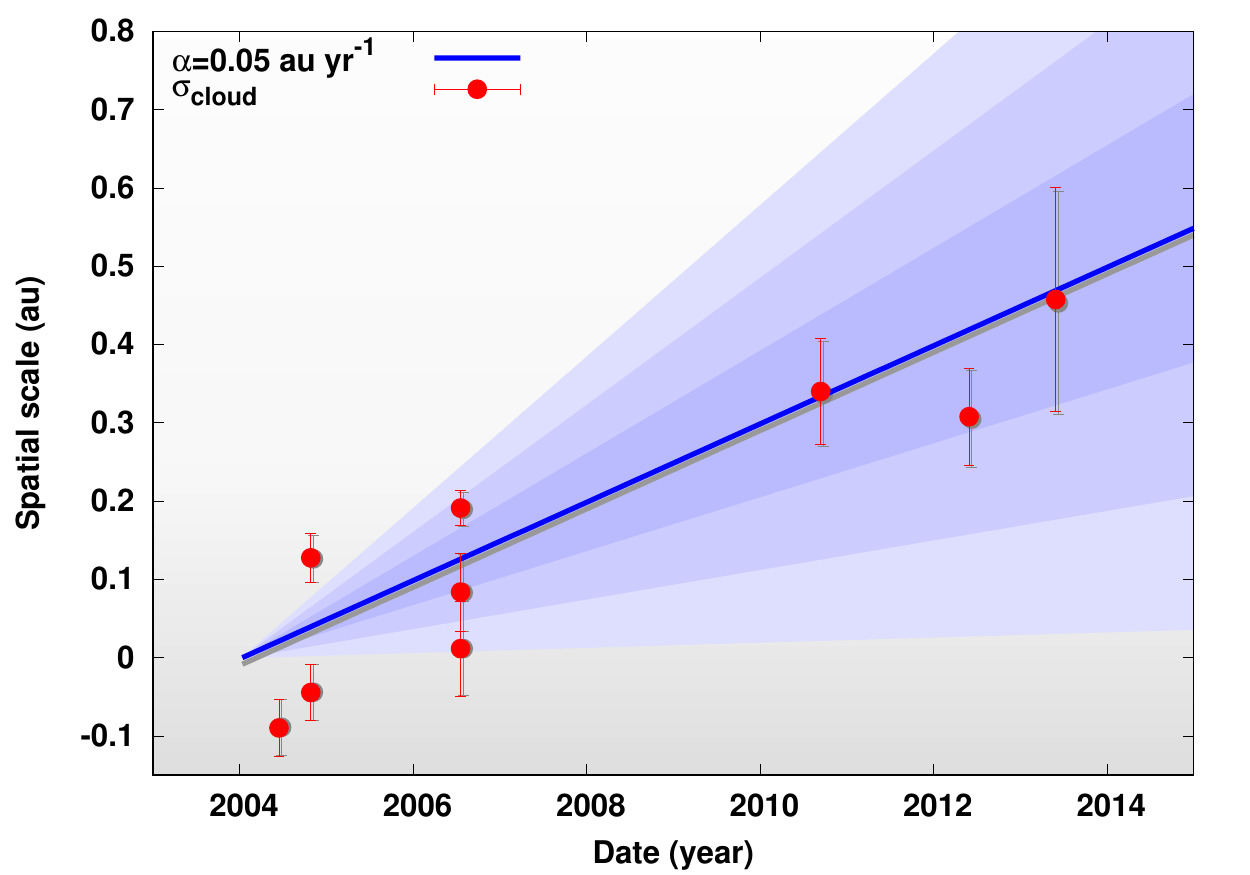}
\caption{The spatial evolution of the observed dust cloud and the best fitting slope with a size increase of  
$0.05\pm0.016$ au$\cdot$yr$^{-1}$, which is approximately 6.2\% of the Keplerian velocity at the collisional location. 
We converted the observed radial profiles to spatial scales by subtracting the wavelength-dependent Gaussian of 
the {\it HST} PSFs from the observed profile widths and multiplying by the distance to the system. The shaded blue areas
show the 1, 2, and 3$\sigma$ regions of the fitting. The date intercept of the fit (and the estimated collision time) is 
at 2004.03 based on the expansion, only slightly earlier than predicted by the astrometric fits.} \label{fig:sigma}
\end{figure}

An additional constraint on the model is the speed at which the cloud expands. While the original launch velocity (combined with 
radiative forces for the dust particles) will determine the average central line of the trajectory, the collisional fragments 
will also diverge radially from the central line (trajectory of the COM) due to the explosive nature of the event that 
produced them. In Fig.\ \ref{fig:radial}, we show the observed evolution of the radius of the dust cloud. The observed size 
depends on the instrumental PSF, the pixel scale, the reduction algorithm, and observatory guiding artifacts. We estimate and correct for these in our analysis in Fig.\ \ref{fig:sigma}. In 10 years the cloud has expanded considerably, to the point where it is larger than the HST PSF (i.e., partially resolved).

These measurements can be used to place limits on the collisional velocity that produced the cloud. Laboratory measurements have 
shown that the speed of daughter particles in collisions, relative to a stationary COM, are roughly identical 
to the relative impact velocities \citep{waza85}. Due to similar orbital paths between colliding bodies, the average collisional 
velocities in disks are around 7.5\% of the orbital velocity \citep{lissauer93}. It is unlikely that either of the colliding 
bodies was originally on a high-speed unbound orbit, so the daughter particles will likely acquire an additional velocity around 
5-10\% of a typical Keplerian velocity at the location of the collision, in a random direction relative to the COM. Some variation 
in this fraction is expected, as the collision did not happen in a disk. The inherited launch velocity of the cloud will then likely project 
onto a bound orbit as well. Gravitational forces will be dominant for the larger and undetected bodies, so they will remain in bound orbits and undergo
Keplerian shear due to the small variations in launch velocity. However, the smallest particles (observed by {\it HST}) will experience additional 
radial acceleration due to radiative forces and be deflected onto unbound orbits. The expansion of the cloud is dominated by
the random velocities acquired during the collision for this source. 

In Fig.\ \ref{fig:sigma}, we illustrate the expansion of the cloud. From each measured radial value, we subtract the wavelength-dependent
standard deviation of the approximating Gaussian ($\sigma$) of the {\it HST} PSF ($\sigma_{\rm STIS} = 0\mbox{$.\!\!^{\prime\prime}$}033$, 
$\sigma_{\rm F435W} = 0\mbox{$.\!\!^{\prime\prime}$}019$, 
$\sigma_{\rm F606W} = 0\mbox{$.\!\!^{\prime\prime}$}026$, and $\sigma_{\rm F814W} = 0\mbox{$.\!\!^{\prime\prime}$}035$). 
These values include PSF broadening due to the Lyot stop of the coronagraphs and also guiding drift. The instrumental PSF broadens due to 
the pupil stops, which restrict the {\it HST} aperture diameter to 88.5\% and 85.5\% of its nominal value for the ACS and STIS detectors, respectively.
Additionally, guiding {\it HST} is difficult when observing Fomalhaut and is typically done using a single guide star, possibly resulting in tracking drifts. 
We measured a per-orbit drift of 0\mbox{$.\!\!^{\prime\prime}$}015, which corresponds to 0\mbox{$.\!\!^{\prime\prime}$}0015 per exposure 
(2.9\% of a pixel, i.e., it is negligible). The 2004 ACS images 
have widths approximately equal to the theoretical values, i.e.\ the cloud is still a point source and therefore the collision likely 
happened close in time to the first observations. However, the dust cloud is visibly extended in the STIS images where it is observed 
(2010, 2012, and 2013). The spatially extended nature of the source in the 2010 and 2012 images was also already noted by \cite{kalas13}. 
They give minor and major axis dimensions; if we take the geometric mean, the agreement with our radial average is within 10\%.
As a third independent confirmation of the extended nature of the Fomalhaut b source in the STIS images, we investigated the
size of point-like background features in the raw data. A visibly slightly elongated peak, possibly the core of a background galaxy, had a 
fitted standard deviation of only 0\mbox{$.\!\!^{\prime\prime}$}042 $\pm$  0\mbox{$.\!\!^{\prime\prime}$}008, more than two times 
smaller than that of Fomalhaut b. The extended Fomalhaut b source is well sampled in the STIS images at their respective epochs; 
therefore, pixel-scale level artifacts do not need to be considered. Following fitting, the expansion of the cloud follows a slope of $0.050\pm0.016$ 
au$\cdot$yr$^{-1}$, which corresponds to an orbital velocity of the colliding bodies equal to 0.5$\pm$0.08 to 1.$\pm$0.16 au$\cdot$yr$^{-1}$,
assuming a collisional velocity 10 to 5\% of the orbital velocity, respectively. The observed expansion of the cloud, therefore, places the 
following constraints on the collision: 
1) The collisional event happened close in time to the first observations in 2004, 2) the colliding bodies had velocities up to $\sim$ 1.2
au$\cdot$yr$^{-1}$, 3) the final launch velocity of the cloud cannot be much higher than this value either. The best fitting launch configuration 
that also conforms to these constraints has the following values: ${\rm x} = 91.23$ au, ${\rm y} = -55.97$ au, ${\rm z} = -9.82$ au, 
${\rm vx} = 0.719$ au$\cdot$yr$^{-1}$, ${\rm vy} = 0.783$ au$\cdot$yr$^{-1}$, and ${\rm vz} = -0.232$ au$\cdot$yr$^{-1}$, yielding a 
reduced $\chi^2=2.48$ and a collisional event that happened 39 days prior to the first observations. This result is in good agreement with the 
collisional time of 2004.03 $\pm$ 0.91, predicted by the expansion. This launch position and velocity will result in a bound orbit of 
${\rm a} = 334.4$ au, ${\rm e} = 0.695$, and $\iota = 12.6^{\circ}$ for the larger undetected fragments of the collision. The smaller 
dust particles, as discussed prior, will be ejected due to the influence of radiative forces. In Fig.\ \ref{fig:astrometry}, we plot 
their trajectory (assuming $\beta=10$), with positions marked at each observation date.

With this best-fitting initial location and launch velocity we used {\tt DiskDyn} to model the dynamical evolution of 
the dust cloud. Due to its larger uncertainty and importance in determining the decrease in surface brightness of the cloud,
the SD of the radial expansion velocities of the dust particles was kept as a variable. A handful of 
models with $\sigma_{\rm v}$ between 0.05 and 0.15 au$\cdot$yr$^{-1}$ were explored. We generated $10^6$ dust particle tracers 
with a size distribution slope of $\alpha = -3.65$. This value is found to be typical of debris disks \citep{gaspar12}, and it is 
consistent with the value of $\alpha = -3.50$ found for particles $>$ 100 $\mu$m (to make the slope not influenced by PR drag) in 
zodiacal dust bands formed in the recent collisional breakup of asteroids \citep{nesv2006}. Slightly shallower 
but similar ($\alpha \sim -3$ to $-3.3$) slopes were found for the size distribution of dust sublimating off comets and in disrupted
asteroids \citep{moreno16b, moreno16c, moreno16a}. The minimum particle size was 0.1 $\mu$m (the largest size is irrelevant, 
as long as all sizes that are efficient at scattering optical light are included). The total dust mass was calculated from 
scaling our calculations to the observations.

The {\tt DiskDyn} dust cloud model was evolved on a GeForce GTX Titan Black GPU, using 0.01 year time steps. For the central 
star, we assumed Fomalhaut's physical parameters (${\rm L} = 15.36~{\rm L}_{\odot}$, ${\rm R} = 1.8~{\rm R}_{\odot}$, 
${\rm M} = 1.92~{\rm M}_{\odot}$, ${\rm d} = 7.7~{\rm pc}$). Dust particles from radii of 0.1 $\mu$m to 1 mm were 
generated in 290 logarithmically spaced size steps, with optical constants for astrosilicates \citep{weingartner01}. 
We performed Mie scattering calculations, taking into account the scattering angles between Fomalhaut, the dust grains, 
and our viewing angle of the system. The results of our modeling are also shown in Fig.\ \ref{fig:radial}. The spatial 
dissipation of the cloud as it propagates on its trajectory is clear, as well as a global trajectory that is increasingly 
radial. 

\subsection*{Morphology and Brightness of the Modeled Dust Cloud}

As the dust particles move further from Fomalhaut, their individual scattered light fluxes drop; the spherical expansion 
of the cloud results in its overall surface brightness dropping as well. To simulate the observations, we convolved the 
model images with their appropriate ACS and STIS PSFs. The model images shown in Fig.\ \ref{fig:radial} take into account 
the convolution and pixel sizes. We measured the fluxes in the model images using the same aperture radius and aperture 
corrections as we did for the data and then finally scaled the model to the observations.

\begin{table}[t]
\centering
\caption{Modeled position and brightness of the Fomalhaut b object\label{tab:model}}
\begin{tabular}{llrrr}
Date    & Instrument &  $\Delta$R.A.\ ($^{\prime\prime}$) & $\Delta\delta$ ($^{\prime\prime}$) & F (Jy)\\
\midrule
2004/Jun & ACS (F814W)   & -8.583                  & 9.153                                  & 3.925e-7 \\
\hline
2004/Oct & ACS (F606W)   & \multirow{2}{*}{-8.593} & \multirow{2}{*}{9.187} & 5.605e-7 \\
2004/Oct & ACS (F814W)   &                         &                                                & 3.896e-7 \\
\hline
2006/Jul & ACS (F435W)   & \multirow{3}{*}{-8.648} & \multirow{3}{*}{9.356} & 4.684e-7 \\
2006/Jul & ACS (F606W)   &                         &                                                & 4.811e-7 \\
2006/Jul & ACS (F814W)   &                         &                                                & 3.208e-7 \\
\hline
2010/Sep & STIS                 & -8.836                  & 9.822                  & 3.907e-7 \\
\hline
2012/May & STIS                 & -8.936                  & 10.040                 & 2.562e-7 \\
\hline
2013/May & STIS                 & -9.000                  & 10.174                 & 1.885e-7 \\
\hline 
2014/Sep & STIS                & -9.093                  & 10.360                 & 1.574e-7 \\
\bottomrule
\end{tabular}
\begin{flushleft}
\addtabletext{The fluxes given were measured the same way as the observations, and therefore 
may not yield the total modeled cloud flux. Positions are of the COM.}
\end{flushleft}
\end{table}

The model predictions varied as a function of $\sigma_{\rm v}$, the SD of the 
additional velocities of the daughter particles. The best fit was at $\sigma_{\rm v} = 0.13 \pm 0.01$ 
au$\cdot$yr$^{-1}$, which is 12\% of the launch velocity and within 3 SD of the 
expansion slope fit. The results of the photometry of the model images, assuming this best fitting
$\sigma_{\rm v}$ value, are shown in Fig.\ \ref{fig:photometry}. The correspondence between the modeled 
observations and the actual photometry is excellent. The PSF convolutions and pixel rebinning, combined 
with the dissipation of the cloud, resulted in a higher peak in the STIS model images, just as observed. 
This is an observational artifact. The overall apparent decrease from 2004 to 2013 is a result of using 
photometry suitable for a point source on a source with increasing extent and decreasing surface brightness. 
Given the PSF subtraction artifacts in the image, photometry optimized to capture the full flux with a 
larger aperture is not possible. Based on the 2014 observations, our model, and the lack of {\it HST}
measurements in subsequent years, future detection of the dust cloud is unlikely.
In Table \ref{tab:model}, we summarize the astrometry and photometry results of our best fitting model. 

\subsection*{Summary of Modeled Dust Cloud Dynamics}
As a result of a catastrophic collision between two massive asteroids on the order of 100 km in radius, a short time prior to the first
set of observations in 2004, a high surface density of dust particles ($\sim 4.8\times10^{22}$ cm$^2$) was produced in the Fomalhaut system.
These dust particles represent the small end of a continuous distribution of fragments, up to tens of km in radius. The fragments continue
on the COM trajectory of the two asteroids, with a velocity of 1.09 au$\cdot$yr$^{-1}$, which is just slightly higher than the
circular Keplerian solution of 0.89 au$\cdot$yr$^{-1}$ at the location of the collision at 107 au. The explosive event disperses the fragments
radially from the COM trajectory, with a variation of $\sigma_v=0.13$ au$\cdot$yr$^{-1}$, which is 12\% of the launch velocity and 15.5\%
of the local circular Keplerian solution. The smallest particles observed with {\it HST} in scattered light experience radiative forces,
resulting in their increasingly radial trajectory.

\subsection*{Probability of Massive Collisions in the Fomalhaut System}

Our photometry model indicates a dust cloud mass of $\sim 1.65\times10^{-8}~{\rm M}_{\rm Earth}$, integrating up to 
1 mm in radius [assuming a density of 3.5 g$\cdot$cm$^{-3}$ appropriate for the likely composition \citep{ball2016}]. 
This equals a scattering surface area of $4.8\times10^{22}$ cm$^{-2}$, assuming a simple $\pi r^2$ surface 
area of scattering. The exact total dust mass depends on the size distribution slope, which is why we also
provide the scattering surface. Producing this amount of dust requires collisions of much more massive bodies.
The required mass and size do not depend strongly on the nature of these bodies. Consolidated objects can fragment 
efficiently \citep{Elmir2019} with fragments extending up to 10 to 50 m 
in size \citep{durda2007, jutzi2019}. Loosely bound rubble piles convey impact energy inefficiently 
\citep{ben2018} and have boulders up to similar sizes \citep{mich2008,mich2019} 
that will survive impact. Integrating the distributions from the smallest up to 
the largest fragments to determine total masses yields similar results for both cases. 

\begin{figure}[!t]
\centering
\includegraphics[width=0.99\linewidth]{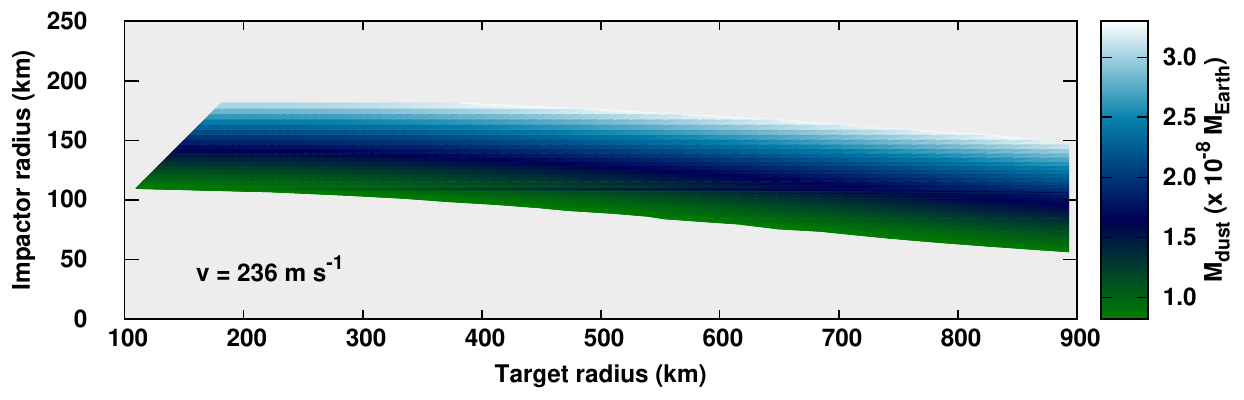}
\caption{The amount of dust produced up to radii of 1 mm as a function of the radii of the target and the impactor, assuming a 
collisional velocity of 236 m s$^{-1}$. Domains producing 0.5 to 2x the observed dust mass are shown.}
\label{fig:mdust}
\end{figure}

How frequently would major collisions occur in the Fomalhaut system? We illustrate the answer by considering  
collisional outcomes and mutual collisional probabilities in a distribution of planetesimals. 
The total dust mass produced in the distribution of fragments in a single collision is
\begin{equation}
    M_{\rm fr}(\mu, M) = \int_0^{Y(\mu,M)}A(\mu, M) m^{-\gamma+1}{\rm d}m\;,
    \label{eq:1}
\end{equation}
where $\mu$ is the mass of the smaller, $M$ is the mass of the larger object partaking in the collisional 
event that produced the distribution, and $\gamma$ is the mass-distribution slope ($\alpha=3\gamma-2$). 
The variable $Y$ gives the mass of the largest fragment in the continuous distribution (i.e.\ it is the second largest fragment 
overall) and $A$ is its scaling factor. 
The redistributed mass also equals the masses of the two colliding bodies, minus the overall largest 
fragment produced ($X$), which is not part of the continuous fragment distribution,
\begin{equation}
    M_{\rm fr}(\mu, M) = \mu + M - X(\mu, M)\;,
    \label{eq:2}
\end{equation}
allowing the scaling factor to be calculated. The values of $X$ and $Y$ are defined by the tensile strength 
of the bodies colliding, their sizes (masses), and their relative velocity. 
Depending on those factors, the collision will either be catastrophic (completely destroying both bodies) or only erosive. 
Our paper \citep{gaspar12} discusses these collisional outcomes in more detail. For tensile strength, we 
assume the curve of \cite{benz99}, traditionally used in debris disk cascade calculations. For the collisional speed, the 
expansion of the dust cloud provides a good estimate, as it will be roughly equal to the relative velocity of colliding bodies. 
The size of the cloud  yields a collisional speed of 0.05 au$\cdot$yr$^{-1}$, while the photometric modeling results in 0.13 au yr$^{-1}$. These correspond 
to 236 and 616 m s$^{-1}$, respectively. In Fig.\ \ref{fig:mdust}, we show the dust mass (up to radii of 1 mm) produced in collisions
between various target-impactor sizes at a collisional speed of 236 m s$^{-1}$, in the domain where the produced dust mass is half 
to twice as much as is observed in the dust cloud. The smallest target capable of producing so much dust is 109 km in radius 
(when hit by a similar-size impactor); however, in erosive collisions, a 56 km body impacting a 893 km object can also generate this 
amount of dust. At the faster impact velocity, the catastrophic outcome remains approximately the same, while the erosive limit 
drops the impactor radius to 47 km. The escape velocity from the largest fragment in the catastrophic collision 
is around $\sim$ 137 m s$^{-1}$, suggesting a high impact velocity. However, the uncertainty in the expansion velocity is 
large and a larger impact velocity would not change our results. 

The rate of collisions between planetesimals that could produce the required amount of dust depends on 
their number density and interaction velocity. Since Fomalhaut b is located near the inner edge of the 
Fomalhaut debris belt, we assume the density of parent bodies at the collisional location to be similar to that 
in the belt. We calculate the belt mass and density from its spectral energy distribution and thermal surface brightness, 
yielding a total of 26.15 M$_{\rm Earth}$, assuming a largest body in the distribution has a diameter of 2000 km and that the belt material 
has a bulk density of 3.5 g cm$^{-3}$. The  timescale for collisional events producing the required amount of dust 
can be estimated by integrating the differential rate of collisions in the projectile and target mass space 
highlighted in Fig.\ \ref{fig:mdust}. We integrate the domain that produces dust half to double of the amount 
observed, yielding timescales  of $\sim$ 0.59 and $\sim$ 0.15 million years for the 236 m s$^{-1}$ and 616 m s$^{-1}$ 
collisional velocities, respectively. 

Fomalhaut b was above the detection level for only a decade, making it unlikely to have arisen in a dynamically cold quiescent planetary system given the 
timescales just derived. Even if the dust production is more efficient than we have assumed, i.e. the slope of the size-frequency-distribution is steeper than 
-3.65, the rarity of this object is shown by the lack of similar sources among {\it HST} images of exoplanets and debris disks \citep{hughes18} compared with 
the near-infrared rate of true exoplanet detections \citep[e.g.,][]{wagner19}.

Dynamical instability from planetary migration can stir planetesimal populations and increase their collision rates. 
It is the most likely explanation for events like Fomalhaut b, and was also invoked by \cite{lawler15}. 
Such dynamical activity has been modeled in detail to explain the influence of planetary migration on the evolution of the 
Solar System \citep{gomes05,strom2005,levison2011}, leading to elevated planetesimal collision rates as reflected, 
for example, in the Heavy Bombardment  indicated by the impact rate on the Moon \citep[e.g.,][]{malhotra01,wu11,Davies14,morbidelli18}.

\section*{Summary}

The planetary nature of Fomalhaut b has been a mystery ever since its detection over a decade ago. In this paper, 
we present previously unpublished measurements of this object, and also re-reduce all archival data to present a 
coherent analysis that shows its behavior over a decade. 

We find that the source has grown in extent since its discovery. We use updated astrometry and orbital solutions,
finding its motion is consistent with radial (escaping) motion. To explain these new observations, we model
Fomalhaut b as an expanding dust cloud, containing copious amounts of dust produced in a massive planetesimal collision.
Our model produces a light curve, angular extent, and orbital motion consistent with the observations spanning a decade.
While Fomalhaut b is not likely to be a directly imaged exoplanet, it is probably the first super catastrophic 
planetesimal collision observed in an exoplanetary system! Production of this amount of dust through planetesimal 
collisions in dynamically quiescent systems should be very rare. The rate of such events would be increased 
substantially if hypothetical planets around Fomalhaut are undergoing orbital migration, resulting in a dynamically 
active population of planetesimals. 

\section*{Data availability}

The raw observational data can be downloaded from the MAST website maintained by the STScI. Our modeling code, 
DiskDyn, can be downloaded from \href{https://github. com/merope82/DiskDyn}{https://github. com/merope82/DiskDyn}.

\acknow{We are grateful for the hardware donation from Nvidia. We thank Paul Kalas for planning and obtaining the observations described here and our referees for providing helpful comments throughout the multiround review process. We also sincerely appreciate the postacceptance in-depth review of an independent {\it HST}/STIS expert. This work has been supported by NASA Exoplanets Research Program Grant 80NSSC20K0268 and NASA Grant 80NSSC18K0555.}

\showacknow{}


\end{document}